\documentclass[sigconf, screen, nonacm]{acmart}
\AtBeginDocument{%
  \providecommand\BibTeX{{%
    \normalfont B\kern-0.5em{\scshape i\kern-0.25em b}\kern-0.8em\TeX}}}

\setcopyright{acmcopyright}
\copyrightyear{2021}
\acmYear{2021}
\acmDOI{10.1145/1122445.1122456}

\acmConference[CHI '21]{CHI '21: Evaluating User Experiences in Mixed Reality}{May 07, 2021}{Yokohama, Japan}

\usepackage{subcaption}  
\usepackage{tabu}  

\begin{document}

\title{"Let's Make A Story": Measuring MR Child Engagement}

\author{Duotun Wang}
\email{duotun@umd.edu}
\affiliation{%
  \institution{University of Maryland}
  \city{College Park}
  \state{MD}
  \country{USA}
}

\author{Jennifer Healey}
\email{jehealey@adobe.com}
\affiliation{%
  \institution{Adobe Research}
  \city{San Jose}
  \state{CA}
  \country{USA}
}

\author{Jing Qian}
\email{jing_qian@brown.edu}
\affiliation{%
  \institution{Brown University}
  \city{Providence}
  \state{RI}
  \country{USA}
}

\author{Curtis Wigington}
\email{wigingto@adobe.com}
\affiliation{%
  \institution{Adobe Research}
  \city{San Jose}
  \state{CA}
  \country{USA}
}

\author{Tong Sun}
\email{tsun@adobe.com}
\affiliation{%
  \institution{Adobe Research}
  \city{San Jose}
  \state{CA}
  \country{USA}
}
\author{Huaishu Peng}
\email{huaishu@cs.umd.edu}
\affiliation{%
  \institution{University of Maryland}
  \city{College Park}
  \state{MD}
  \country{USA}
}



\begin{abstract}
We present the result of a pilot study measuring child engagement with the "Let's Make A Story" system, a novel mixed reality (MR) collaborative storytelling system designed for grandparents and grandchildren.  We compare our MR experience against an equivalent paper story experience.  The goal of our pilot was to test the system with actual child users and assess the goodness of using metrics of time, user generated story content and facial expression analysis as metrics of child engagement.  We find that multiple confounding variables make these metrics problematic including attribution of engagement time, spontaneous non-story related conversation and having the child's full forward face continuously in view during the story. We present our platform and experiences and our finding that the strongest metric was user comments in the post-experiential interview.
\end{abstract}




\begin{teaserfigure}
  \includegraphics[width=\textwidth]{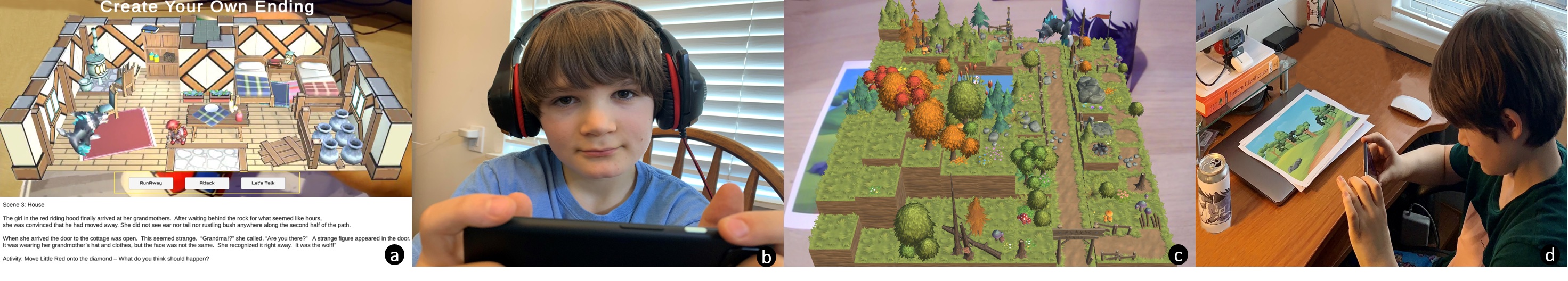}
  \caption{"Let's Make A Story" system makes it possible for the grandparent and grandchild experience the story simultaneously but with different views and roles: a) and b) grandparents could control the story environment and see their grandchild's face and talk to them during MR story play; c) and d) grandchild could manipulate the can-controlled character and hear their grandparent's voice }
  \label{fig:teaser}
\end{teaserfigure}

\maketitle

\section{Introduction}
In a world where people are often separated by distance, people crave meaningful experiences that can happen remotely.  The "Let's Make A Story" system is designed to create an engaging collaborative experience for physically distant grandparents and grandchildren.  We present our system and our insights from a set of three experiences with children ages 6,7 and 9.  Our application builds on paradigm of storytelling, where an adult will read a story to a child and share a bonding experience. We address the challenge that it is difficult to engage children's attention for long periods of time over video calls \cite{StoryVisit} and leverage the fact that children today are often frequently engaged by phone video games.  We designed a story sharing system that addresses both the needs of grandparents and grandchildren.  For the grandparents, we created a desktop application that includes the words of the story and various controls for the story environment.  The grandparents control the pace of the story and actions that are necessary for the story to advance.  Grandchildren are given a phone based application that resembles a game and a manipulative, a can wrapped in an MR triggering image, to control the protagonist in the story.  Grandparents see and and control the game and also see their grandchild's face while children are game focused while hearing their grandparents Voice.  The twin experiences are shown in Figure  ~\ref{fig:teaser}.  While our initial plan was to measuring engagement by story time, amount of supplemental story generated by the user and facial expression analysis, we found that there are many practical considerations that need to be considered to make these viable metrics.  We instead find that user response in the post-hoc interview is highly informative.

\section{Related Work}
Video conferencing has been found to be more highly engaging than phone calls when it comes to connecting grandparents and grandchilden, as children are often at a loss when it comes to sustaining conversation without expressing themselves physically  \cite{LoveCloset}.  Interacting with grandchildren was found to be the "the primary, if not sole, motivation for video" \cite{LoveCloset} for grandparents who would prefer the phone.   Videochat allows kids to be more engaged\cite{VideoPlay} and  allows them to assert their own participation more easily by putting something (an object, their body) in front of the camera rather than needing to find words to share\cite{GrandparentGrandchild}.  

Beyond videochat alone, systems such as Family Story Play system\cite{Raffle10} and StoryVisit\cite{StoryVisit} allow videoconferencing to be augmented with remote storytelling.  While StoryVisit uses a shared ebook, Family Story Play wuses a paper book with an augmented frame that plays videos of a popular television character at different parts of the story.  Both these studies found that augmented storytelling increased child engagement using time in story as a metric. 

The problem of remote grandchild-grandparent engagement still has not been solved \cite{GrandparentGrandchild, VirtuallyCOVID, VideoPlay}. Many of the reasons cited in prior work include the disparity in the levels of technical literacy and comfort between the generations\cite{VideoPlay, StoryVisit, LoveCloset}.  We designed our mixed reality experience with these concerns in mind, providing an easy to use desktop interface for grandparents and a fun mobile phone based MR experience with tangibles for grandchildren. 
\section{System Overview and Interaction}
\label{section: system}
\begin{figure}
\includegraphics[width = \columnwidth]{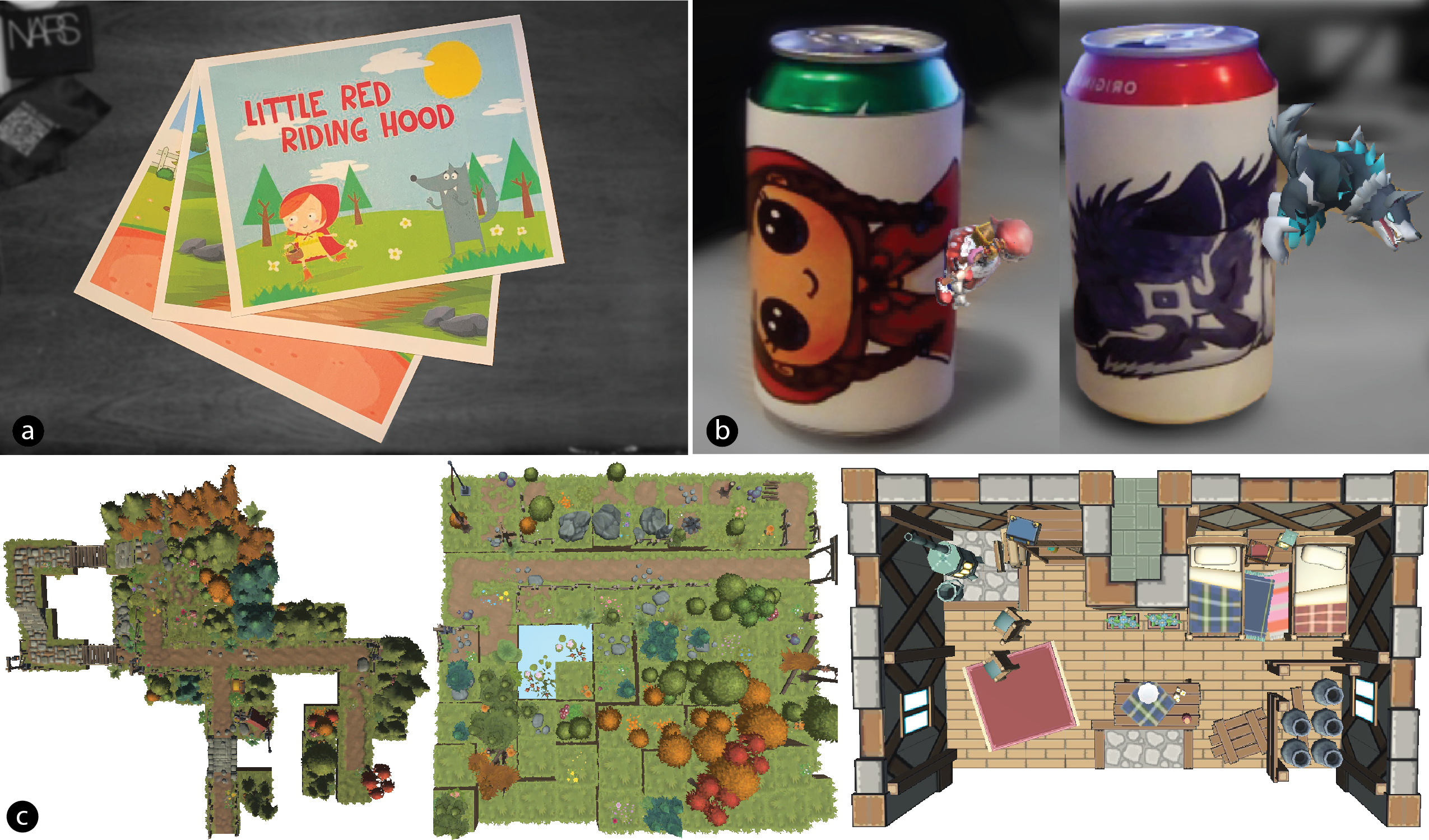}
\caption{a) Printed book pages were used for generating c) interactive virtual background for the MR Story Play; b) Cans were used as MR Characters (little red and wolf). Book images were redesigned from Behance \protect \footnotemark. 3D Models were from Unity Asset Store \protect \footnotemark .}
\label{fig:ARBackground}
\end{figure}

Our system, inspired by prior work\cite{Raffle10, StoryVisit, hettiarachchi2016annexing}, consists of a desktop program for the grandparents (Figure ~\ref{fig:teaser}a) and a mobile phone mixed reality application (Figure ~\ref{fig:teaser}c). The grandparent reads the text of the story and controls story-related events (dropping apples, controlling endings).  The child  generates 3 MR scenes (Figure ~\ref{fig:ARBackground}c) using a 3-page printed paper book (Figure ~\ref{fig:ARBackground}a) and controls a manipulative character using a 12oz can wrapped in an MR-activating sheath (Figure ~\ref{fig:ARBackground}b). The grandparent begins the application using the desktop app to invite the child to "join". The grandparent has reads the text of the story and controls the story-related events. The child uses the can to make the MR character interact with the virtual background displayed on the phone. Each page ends with MR activity initiated by the grandparent and completed by the child.

\begin{figure}
\includegraphics[width = \columnwidth]{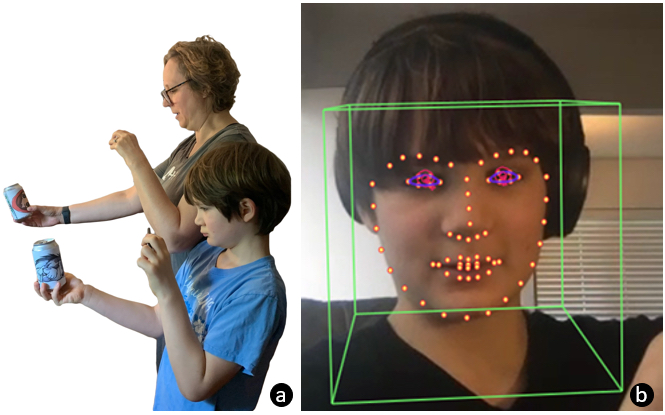}
\caption{a) Children with short arm has difficulties positioning the phone camera in related to the can. b) Using OpenFace to measure engagement.}
\label{fig:Challenges}
\end{figure}

\footnotetext[1]{House Interiors:
\url{https://assetstore.unity.com/packages/3d/environments/fantasy/retro-dungeons-house-interiors-170705}\\
Isometric Pack 3D:
\url{https://assetstore.unity.com/packages/3d/environments/fantasy/isome-tric-pack-3d-62262}
}
\footnotetext[2]{Little Red Book Cover: 
\url{https://www.behance.net/gallery/9151145/Little-Red-Riding-Hood-(Book-Covers)}}

\footnotetext[3]{Werewolf Cute Series: \url{https://assetstore.unity.com/packages/3d/characters/creatures/werewolf-cute-series-177868}}

\section{User Study}
We conducted a pilot study with three participants, two boys aged 6 and 9 and one girl aged 7. All participation was remote.  An Android APK for the MR experience was customized for each participant's phone as camera resolution is an important factor.  Each participant's parents were given instructions on how to assemble all paper materials which were printed by the participants.  This pilot was focused on measuring child engagement, so for convenience the role of the grandparent was played by one of the experimenters.  Each child participant read two stories, one paper based story as a benchmark and one story with the MR game. Both stories were versions "Little Red Riding Hood," one version told from Little Red's perspective and the other from the Wolf's. The MR background, shown in Figure ~\ref{fig:ARBackground}c was the same for both stories, but the can based characters, shown in Figure ~\ref{fig:ARBackground}b and active elements were protagonist specific (e.g., apples vs. bunnies, different paths through the forests, different antagonist and action choices in final scene). 

The grandparent leads the interaction and the child follows along. For the paper story interactions include answering questions about the story visuals and lifting paper flaps to find a hidden items. In the MR experience, activities include making the MR background appear and summon the character with the can.  The child is free to play with the character while the grandparent reads the text.  Each page ends with a MR activity like "catch the bunnies", "follow the path" or "choose an action."  At the end of the third page of both stories the child is asked to create an ending.  Children and their parents (if present) are briefly asked about their experiences at the end of the session.

\section{Measurement Challenges} 
Our initial hypothesis was that, similar to prior studies,  we use time as an engagement metric.  We also wanted to explore measuring user generated content (the story ending) and facial expressions as potential engagement metrics.  During the pilot study we encountered several challenges that caused us to reconsider how best to measure engagement, described below.
\subsection{Time Metric}
Interaction time is the most common metric of engagement. The amount of time an experience take, however, can reflect more than just the participant's interest.  For example, we found that children often took a lot of time to get the MR can-controlled character to appear in the scene.  Factors including lack of adequate lighting, children occluding part of the overlay with their fingers and children's shorter arms lengths illustrated in Figure ~\ref{fig:Challenges}a all seemed to contribute to the difficulty. In two of the three cases, simultaneous can and phone manipulation was too difficult and the strategy of placing the can on the table and moving the phone to move the character was adopted (shown in Figure ~\ref{fig:teaser}).  The paper stories also had technical difficulties where sometimes pieces were lost or not assembled properly.  These technical difficulties are confounds to engagement in the time metric. Our initial method to refine interaction time is to perform prepossessing of recorded remote calls by removing the excessive time not used for the story play experience. We also designed a trail session to help children to become familiar with MR can-controller characters before the story play, aiming to reduce the unexpected time elapsed during MR story.

\subsection{Participant Generated Story}
Another metric we had considered was user generated conversation. Throughout the story the grandchild was given opportunities to interact with the grandparent. We found that in the programmed interaction none of the three children elaborated on the story beyond simple answers to the questions. In the final scene, where the children need to make up an ending children's responses varied, but almost all of them were confused at first and needed prompting.  One child was reluctant to say anything in either the paper or the MR ending.  Another child was proud that they knew the actual story of Little Red Riding Hood story form memory at the end of the paper story which was presented first.  In the third case the child was more interested in giving commentary on the difference between the wolf characters in the two stories at the end of the paper story which in this case was presented second. Our pilot testing indicated that the story type  (paper  or MR) was not the primary driver of child commentary.  

\subsection{Facial Expression Analysis}
Another promising engagement metric is facial expression analysis, using software such OpenFace \cite{baltrusaitis2018openface} which can quantify gaze and facial action units used for emotion recognition (Figure ~\ref{fig:Challenges}b). Quantitative metrics such as the duration, frequency, or pattern of emotion involvement can provide more context about the overall experience. For instance, we noticed that children in MR conditions smiled for a longer period of time.

\subsection{Post-Hoc Interviews}
Despite out hope for objective metrics, we believe that the most explicit evidence for engagement came from the subjective post-hoc interviews. When asked about the experience, both the children and the adults co-located with the children expressed a high degree of enthusiasm for the MR experience.  Comments such as "That was epic," "This is awesome" and a follow up text comment "It was SO FUN!" all lead us to believe that the experience was engaging.

\section{Conclusion} 
We believe that our system has the potential to create engaging cross-generational experiences that both grandparents and grandchildren will love.  In our initial pilot we tested the hypothesis that we could measure engagement through story time, user generated story content and facial expression analytics. We found that each of metrics needed to be carefully considered in light of other confounds. Our current best indication of relative engagement has been from the post-hoc user interviews.  In future work we plan to continue to address technical difficulties and to better quantify user subjective experience with an explicit questionnaire.  We had initially wanted to avoid a subjective assessment from children, but from our pilot we found that co-located parents also observed the child's interaction and would be in a position to make an assessment of child's engagement.    


\bibliographystyle{ACM-Reference-Format}
\bibliography{ARStory}

\end{document}